# Spin dependent charge transfer in MoSe$_2$/hBN/Ni hybrid structures

H. Tornatzky[1], C. Robert[1], P. Renucci[1], B. Han[1], T. Blon[1], B. Lassagne[1], G. Ballon[1], Y. Lu[2], K. Watanabe[3], T. Taniguchi[4], B. Urbaszek[1], J.M.J. Lopes[5], X. Marie[1]

*[1]Université de Toulouse, INSA-CNRS-UPS, LPCNO,135 Av. Rangueil, 31077 Toulouse, France*

*[2]Institut Jean Lamour, UMR 7198, CNRS-Université de Lorraine, BP 239, 54011 Nancy, France*

*[3]Research Center for Functional Materials, National Institute for Materials Science, 1-1 Namiki, Tsukuba 305-0044, Japan*

*[4]International Center for Materials Nanoarchitectonics, National Institute for Materials Science, 1-1 Namiki, Tsukuba 305-0044, Japan*

*[5]Paul-Drude-Institut für Festkörperelektronik,Leibniz-Institut im Forschungsverbund Berlin e.V., Hausvogteiplatz 5-7, 10117 Berlin, Germany*

# Abstract

*We present magneto-photoluminescence measurements in a hybrid 2D semiconductor/ ferromagnetic structure consisting of MoSe$_2$/hBN/Ni. When the Nickel layer is magnetized, we observe circularly polarized photoluminescence of the trion peak in MoSe$_2$ monolayer under linearly polarized excitation. This build-up of circular polarization can reach a measured value of about 4% when the magnetization of Ni is saturated perpendicularly to the sample plane, and changes its sign when the magnetization is reversed. The circular polarization decreases when the hBN barrier thickness increases. These results are interpreted in terms of a spin-dependent charge transfer between the MoSe$_2$ monolayer and the Nickel film. The build-up of circular polarization is observed up to 120 K, mainly limited by the trion emission that vanishes with temperature.*

The recent intensive studies of optical[1] and spin/valley properties[2,3,4,5] in two-dimensional (2D) semiconductors based on transition metal dichalcogenides (TMDs) have opened the possibility to propose new optoelectronics and spin/valley-optronics devices where the spin/valley index[2,6] would constitute an additional degree of freedom to carry information. In addition to their particular optical properties due to strong excitonic effects and 2D confinement, direct bandgap TMD monolayers (ML) also present non-equivalent K valleys (with opposite spin properties) in the reciprocal space, due to the lack of inversion symmetry and strong spin-orbit coupling[2,6]. As a consequence, the carrier spin/valley index can be controlled using circularly-polarized light excitation. The manipulation of the spin/valley properties in ML TMDs can also be achieved with an external magnetic field lifting the degeneracy between the two K valleys[7,8]. Nevertheless, large external magnetic fields are generally required to split the energy levels (~0.2 meV/T), making this solution inappropriate in future devices. In this context, a promising approach is to use the proximity effect[9] between a magnetic layer and the TMD ML[10], providing a large effective magnetic field in the semiconductor. The observation of this effect is difficult in hybrid devices based on usual epitaxial semiconductors[9] such as GaAs, Si, CdTe due to the poor quality of the semiconductor/magnetic interfaces[11,12,13]. The inherent atomic scale of the hybrid devices based on TMDs and the absence of dangling bonds at interface make these systems suitable to observe such magnetic proximity effects, and several predictions or observations have been reported, both with magnetic insulators (EuS/WSe$_2$[14]) or ferro(antiferro) magnetic 2D materials (CrI$_3$/WSe$_2$[15,16], CrBr$_3$/MoSe$_2$[17,18], MnO/WSe$_2$[19], Cr$_2$Ge$_2$Te$_6$/WSe$_2$[20], Mn-perovskite/hBN/WSe$_2$[21]). It results in a giant effective Zeeman splitting observed in photoluminescence (PL). Up to now, the operation has been limited to very low temperatures (typically below 60 K), due to the limited Curie temperature of 2D magnetic materials.

On the other hand, the spin-dependent charge transfer between a 2D semiconductor and a magnetic material is also challenging. In terms of electrical spin injection into TMDs, it has been reported in Spin Light Emitting diodes based on GaMnAs/WS$_2$[22] and NiFe/WSe$_2$[23] or in a lateral Co/MgO/MoS$_2$ spin valve system[24]. Concerning spin dependent transfer of carriers from a semiconductor material towards a magnetic material, it was observed in systems based on conventional semiconductors such as GaAs[25] or Ge[26, 27]. Recently, such an effect has been observed for TMD MLs at the interfaces CrI$_3$/WSe$_2$[15], CrBr$_3$/MoSe$_2$[17], and Mn- perovskite/hBN/WSe$_2$[21]. Here again, the effect was evidenced only at low temperature due to the low Curie temperature of the magnetic materials.

An alternative to overcome the temperature limitation of the previous effects is to use ferromagnetic transition metals such as Fe, Co or Ni, with Curie temperatures well above 300 K. In this case, it is necessary to insert a thin insulating barrier such as for example MgO or hBN[28] in order to preserve the semiconductor nature of the TMDs, that would be transformed into metallic state otherwise due to the appearance of interface states within the gap[29]. In this paper, we investigate by magneto-photoluminescence spectroscopy MoSe$_2$ ML/hBN/Ni structures where the hBN/Ni part is grown by molecular beam epitaxy (MBE). We evidence spin-dependent transfer between the MoSe$_2$ ML and the ferromagnetic nickel film, which can be accurately controlled by the thickness of the hBN tunnel barrier.

Figure 1a depicts the structure of our samples: a 400 nm thick slab of Ni is first grown on a MgO(111) substrate by electron beam evaporation. The sample is then transferred to a MBE system where the Ni surface is cleaned (via annealing and Ar$^+$ sputtering) prior to the MBE growth of a ~1 nm thick hexagonal boron nitride (hBN) layer. This procedure provides a clean hBN/Ni interface and protects the Ni surface from oxidation. Details about the Ni surface preparation and the hBN-growth can be found elsewhere[30,31]. An exfoliated ML of MoSe$_2$ is then deposited onto this epilayer using a dry transfer technique[32] and capped by a thin exfoliated high quality hBN flake (thickness around 10 nm)[33]. In order to tune the distance between Ni and MoSe$_2$ layers, we fabricated several similar samples with an additional exfoliated hBN flake below the MoSe$_2$ ML resulting in a total tunnel barrier thickness $d_{hBN}$. The thickness of the additional hBN layers was measured by atomic force microscopy. We also fabricated a control sample on a simple non-magnetic layer: a hBN-encapsulated MoSe$_2$ ML on a SiO$_2$/Si substrate (see the sketch in the insert of Figure 1b).

We then performed magneto-photoluminescence (PL) experiments in a confocal microscope using a HeNe (1.96 eV) laser as the excitation source and a spot size smaller than 1 μm diameter[34]. The average excitation power is 2 μW. The sample is excited with linearly polarized light while we detect the circularly polarized luminescence. A magnetic field is applied perpendicular to the sample (Faraday geometry) using a superconducting coil with a field up to 2T from 4 K to 120 K.

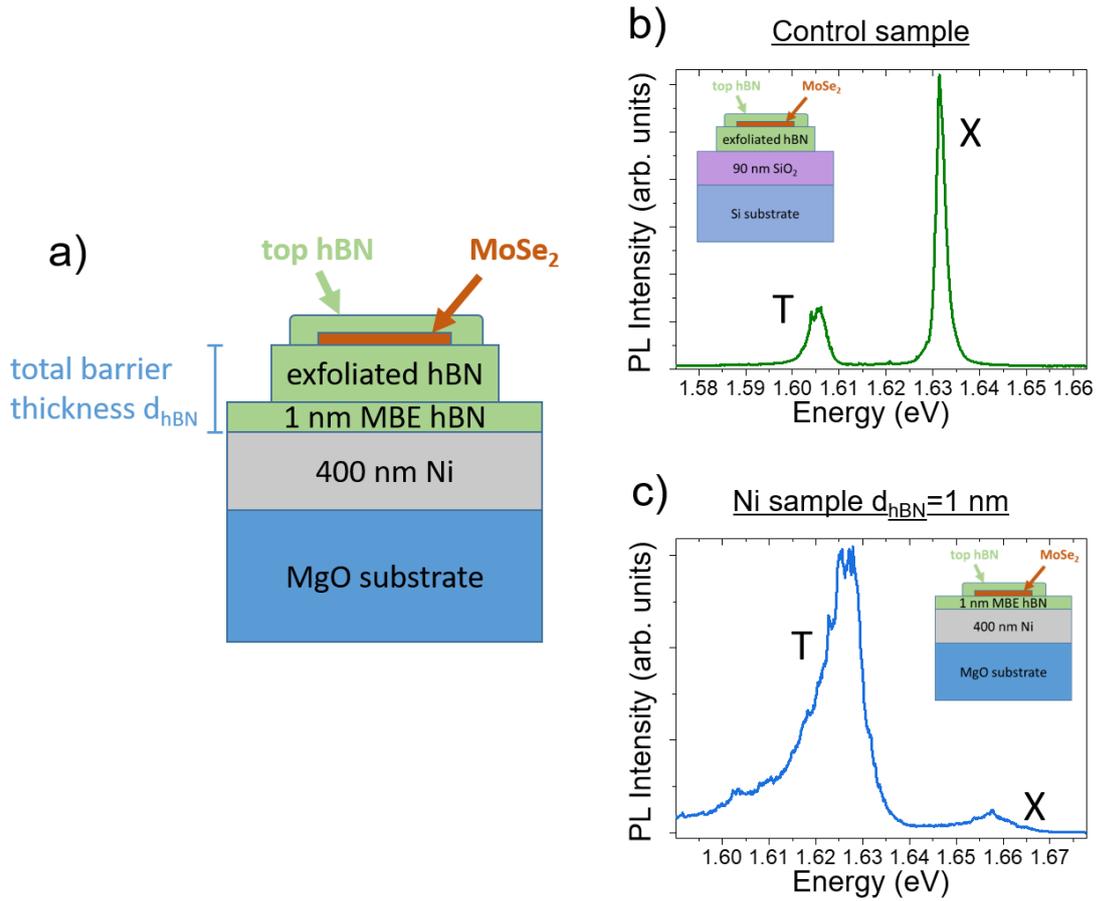

*Figure 1 : a) Schematic of the sample structures. MBE-hBN denotes the hexagonal boron nitride grown directly onto the Ni film by MBE. b) Photoluminescence spectrum of the control sample hBN/MoSe$_2$/hBN/SiO$_2$/Si. c) Photoluminescence spectrum of the Ni sample with the thinnest hBN barrier corresponding to the MBE hBN only (no exfoliated hBN flake). The temperature is 4 K.*

We compare in Figure 1b and Figure 1c the PL spectra taken at 4 K and 0 T for the control sample (without the Ni layer) and the Ni sample with the thinnest hBN barrier ($d_{hBN}$=1 nm). For the control sample, we clearly observe two peaks corresponding to the neutral (X) and charged exciton (trion T) with narrow linewidths (FWHM=2 meV for X), indicating the excellent quality of the structure[35]. In contrast, for the Ni sample with a thin barrier the linewidth of both transitions is broadened and the total PL intensity is quenched by around two orders of magnitude indicating efficient charge and/or energy transfer[36]. We show in the supplementary materials (Figure S1) that the PL intensity increases with the hBN barrier thickness. Interestingly, the ratio of intensities between the trion line and the exciton line is larger for the Ni sample than for the control sample. This can be interpreted as a larger electronic doping of the MoSe$_2$ ML when the Ni layer is close; i.e. a strong indication that electronic transfer occurs at the interface. We show in the supplementary materials (Figure S2) that this ratio decreases rapidly when the hBN barrier thickness increases.

We have then measured the degree of circular polarization $P_c$ of each line following linearly polarized excitation as a function of an external magnetic field. We define $P_c = (I_{\sigma+} - I_{\sigma-})/(I_{\sigma+} + I_{\sigma-})$ where $I_{\sigma+}(I_{\sigma-})$ are the integrated intensities of the right (left) circularly polarized luminescence components. We first show the results for the trion peak in the control sample in Figure 2a. As expected, $P_c$ exhibits a linear behavior which is the consequence of the Zeeman splitting of the bands at $K^+$ and $K^-$ [7,8]. In contrast, the circular polarization degree of the trion peak on the Ni sample with the thinnest hBN barrier deviates from linear behavior with magnetic field, but exhibits a Z-shaped feature (Figure 2b). The Z-shape of $P_c$ is symmetric with respect to the zero field point and exhibits a linear dependence between the positive and negative critical fields $\pm B_c$, at which the slope changes. To investigate this feature, we have subtracted the linear behavior induced by the Zeeman effect by fitting $P_c$ with a piecewise linear function to obtain the slope at $|B| > B_c$ (dashed line in Figure 2b). After subtraction we obtain the additional circular polarization with respect to the one induced by the linear Zeeman effect (Figure 2c) that we note Δ. We clearly see that Δ saturates at about $Δ_{max}=±4\%$ for $|B| > B_c=0.58±0.03$T. Remarkably, the magnetic field dependence of Δ perfectly matches with the out-of-plane magnetization of the Ni film measured by vibrating sample magnetometry (VSM) and shown in Figure 2c by the blue solid line. At zero magnetic field, such a thick Ni film has an in-plane remanent magnetization[36]. The out-of-plane direction is the magnetic hard axis and the corresponding magnetization component follows linearly the applied magnetic field and saturates around 0.58±0.03T, close to the expected value of 0.62T calculated from the Stoner-Wohlfarth model for a Ni infinite film[37,38]. This result proves that the circular polarization of the trion is linked to the out-of-plane magnetization of the Ni film. However, the circular polarization of the PL emission of the neutral exciton is close to linear with the magnetic field and is not related to the magnetization of the Ni layer. (shown in the supplementary materials Figure S3).

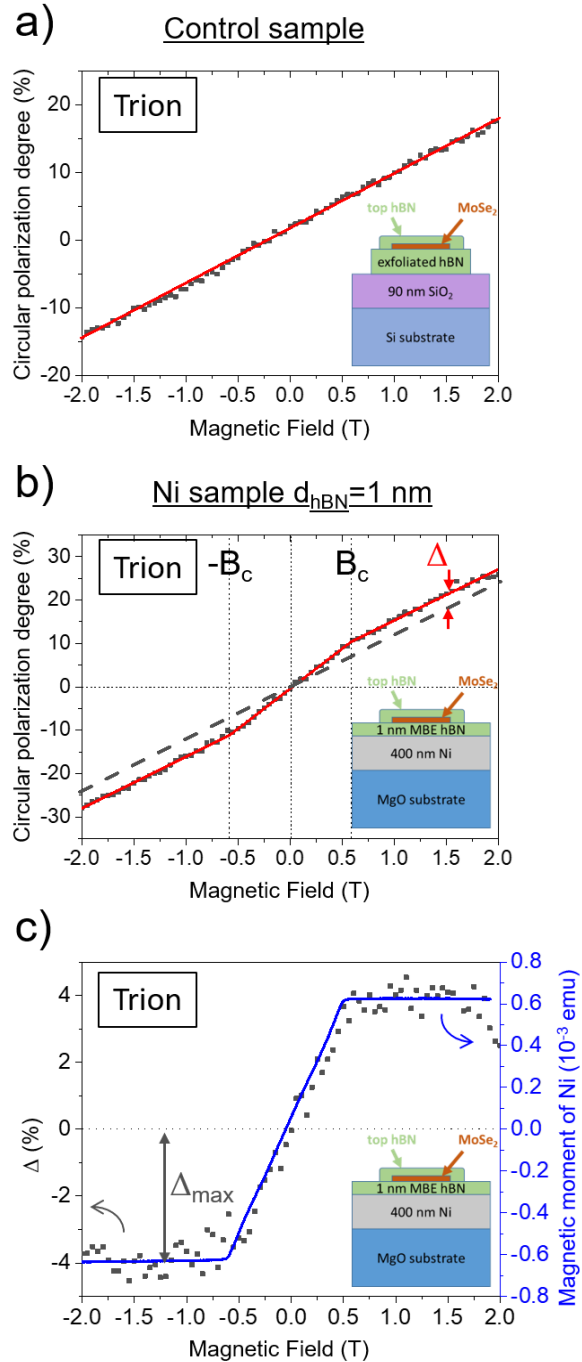

*Figure 2 : Degree of circular polarization as a function of the external magnetic field for a) the trion in the control sample on Si substrate and b) the trion in the Ni sample with the thinnest hBN barrier. c) Build-up of circular polarization Δ of the trion after subtracting the linear behavior induced by the Zeeman splitting shown in b). In addition the magnetization response of the Ni film by VSM is plotted as solid blue line. The temperature is 4 K for both PL and VSM measurements.*

This build-up of circular polarization of the trion could have several origins. First it could be due to Ni magnetic proximity effect based on exchange interaction[28]. Thanks to the estimation of the $P_c$ slope of about 11.5%/T corresponding to the circular polarization due to the usual Zeeman effect, a $\Delta_{max}$ value of 4% would correspond to an effective magnetic field of about 350 mT that would result in a Zeeman splitting of about 70 µeV. In our case, it is not possible to determine such a small Zeeman splitting on spectra due to the width of the trion line. Nevertheless, such an effective magnetic field due to exchange interaction should also lead to equivalent circular polarization due to Zeeman effect on the exciton line[8], and it is not observed in our measurements (see supplementary materials Figure S3). Moreover, as shown in Figure 3a, we measure significant trion polarization for hBN thicknesses up to d~16 nm ; this rules out a possible short-range proximity effect which should be observed only for much shorter barrier thicknesses. Another possible contribution to the PL circular polarization could be linked to the magnetic circular dichroism (MCD) of the Ni layer[39,40], either in terms of reflection of the laser light on Ni that would re-excite the MoSe$_2$ ML, or reflection on Ni of the light emitted by the TMD ML itself, or by more complex cavity effect phenomena where the hBN bottom and top layers would constitute a cavity, terminated by a Ni mirror on the bottom side. However, if MCD was the dominant effect, it should lead to equivalent circular polarization on the trion and exciton lines, and it is not the case experimentally. Moreover, this effect should not vary with temperature in the range 4K-120K, far below the Curie temperature of Ni, contrary to what is observed experimentally for the trion emission (see the insert of Figure 3b). We also rule out a possible interpretation based on stray fields (see supplementary materials 4.).

The remaining interpretation is to consider spin dependent tunneling [27,21]. Carriers could tunnel through hBN into Ni, where the density of final states are different for the two spin species at the Fermi Level, leading to spin dependent tunneling efficiencies, thus unbalanced spin populations in MoSe$_2$ and finally the appearance of circular polarization for the trion. As shown in Figure 1b and 1c, the relative intensity of the trion and exciton peaks is in favor of the trion peak when the MoSe$_2$ ML and the Ni film are close indicating a charge transfer and doping thanks to tunneling through the thin hBN barrier. A further indication of the charge transfer occurring at the interface is observed by analyzing the integrated intensities of the exciton and trion lines as a function of the barrier thickness (shown in the supplementary material Figure S1). A quenching of the PL is visible with the reduction of the barrier thickness. The charge transfer between the MoSe$_2$ ML and the Ni layer is very consistent with the mechanism of spin dependent transfer between the two layers that would result in the build-up of circular polarization in MoSe$_2$. In this case, a vanishingly small Δ on the exciton line (as observed experimentally in supplementary materials Figure S3) is expected due to the much shorter lifetime of the exciton (about 1 ps[41]) with respect to the tunneling time through hBN. In other words, excitons recombine before the tunneling occurs, preventing the effect. The situation is different for trion, whose lifetime is much longer (typically more than ten ps[41]), allowing the build-up of circular polarization.

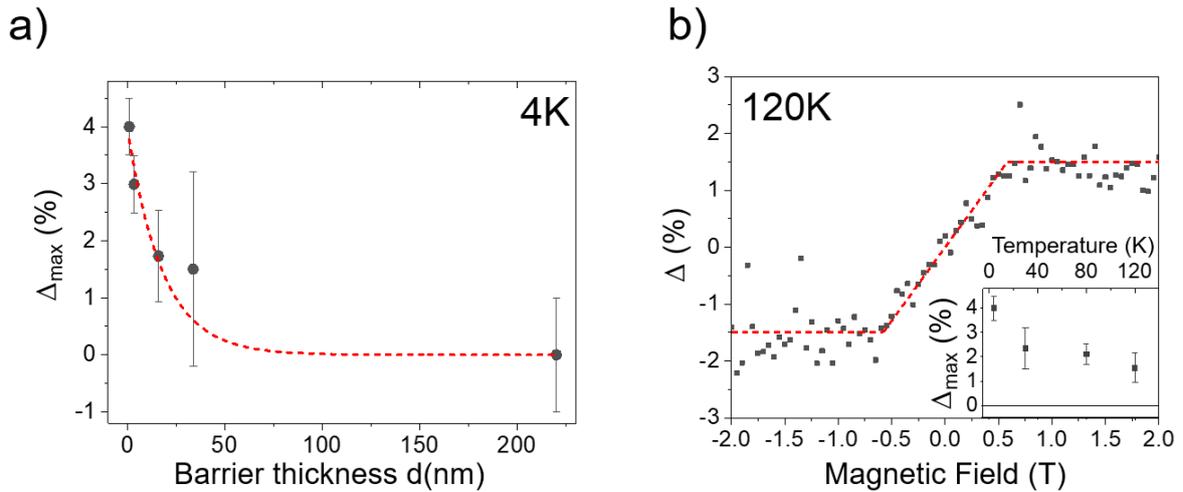

*Figure 3 : a) Variation of Δ$_{max}$ with hBN-barrier thickness at T=4K. b) Build-up of circular polarization Δ of the trion at T=120K for d$_{hBN}$=1nm. The insert shows the variation of Δ$_{max}$ with temperature. The dashed red lines are guides for the eye.*

Finally, we discuss the dependence of the circular polarization Δ$_{max}$ on the temperature. Figure 3b shows the build-up of circular polarization Δ as a function of magnetic field for the sample with d$_{hBN}$=1nm and a temperature of 120K. We clearly observe the same trend than in Figure 2c despite a reduction of Δ$_{max}$ to a value around 1.5%. The measurements as a function of the sample temperature are depicted in the insert. This decreasing trend could be tentatively attributed to a decrease of the carriers spin relaxation time in MoSe$_2$ [42]. These data provide evidence for the robustness against temperature of the spin/valley polarization of a TMD monolayer that is mainly limited by the trion emission that vanishes at 120K (see supplementary materials Figure S4). To the best of our knowledge, this represents the highest temperature for which the influence of a ferromagnet in close proximity to a TMD is observed[12,14,15,17,18].

In conclusion, we have investigated the spin/valley properties of a hybrid MoSe$_2$/hBN/Ni structure by magneto-photoluminescence spectroscopy. Once the Nickel layer is fully magnetized to out-of-plane, we observe the appearance of a circularly polarized photoluminescence of about 4% of the trion peak in MoSe$_2$ monolayer under linearly polarized excitation. This circular polarization follows the out-of-plane magnetization of the Ni layer. The observed circular polarization decreases when the hBN barrier thickness increases. This observed circular polarization build-up may be attributed to spin dependent tunneling between MoSe$_2$ and Ni. Finally, we show that the observed circular polarization is still observable up to 120K, mainly limited by the trion emission that vanishes at this temperature. As expected, this is a key advantage of the MoSe$_2$+hBN/Ni hybrid system compared to previous structures based on 2D ferromagnetic material with low Curie temperature. Note that the use of an epitaxially grown hBN/Ni structure using a conventional transition ferromagnetic metal to build our final hybrid device is a relevant step in view of large area devices, contrary to fully exfoliated small sizes systems. Performing similar experiments on charge tunable devices should allow an additional control of the doping in the MoSe$_2$ ML and could lead to the measurement of the effect at room temperature. In the future spin-dependent transfer generating circular polarization emission in the TMD monolayer could be demonstrated without the need of an external field, by replacing the Ni layer by a magnetic layer presenting perpendicular anisotropy[43–46], such that its magnetization remains perpendicular to the film in the remanent state.

**Acknowledgments:**

This work is supported by the ANR 692 SIZMO2D project (Grant No. ANR-19-CE24-0005), ANR ATOEMS and ANR Magicvalley. Growth of hexagonal boron nitride crystals was supported by the Elemental Strategy Initiative conducted by the MEXT, Japan, Grant Number JPMXP0112101001, JSPS KAKENHI Grant Number JP20H00354 and the CREST(JPMJCR15F3), JST

# Supplementary Materials

1. <u>Role of the hBN barrier thickness on the PL spectra</u>

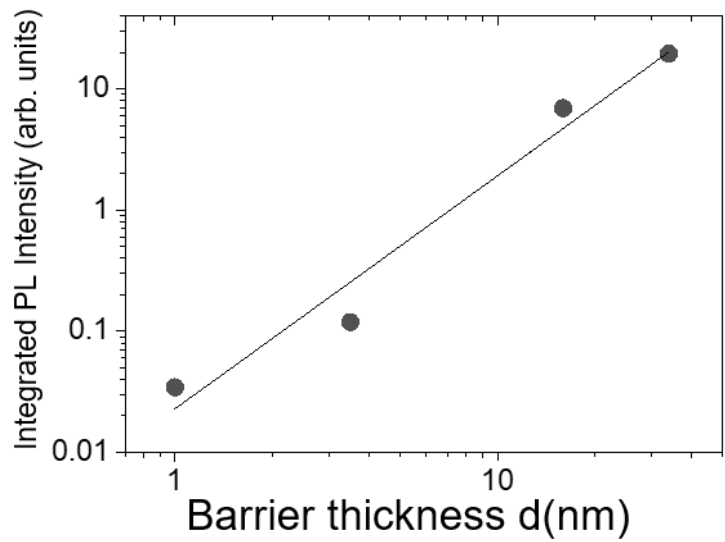

*Figure S1. Integrated PL Intensity as a function of the hBN barrier thickness at T=4 K.*

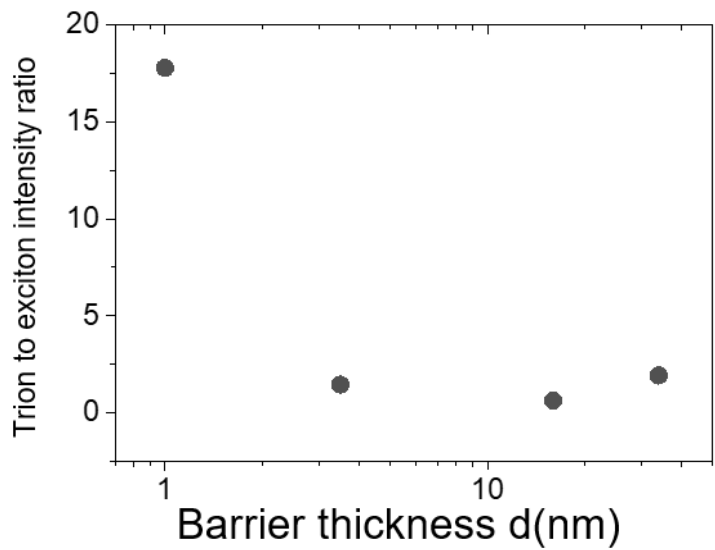

*Figure S2. Trion to exciton PL intensity ratio as a function of the hBN barrier thickness at T=4K.*

2. <u>Circular polarization of the neutral exciton</u>

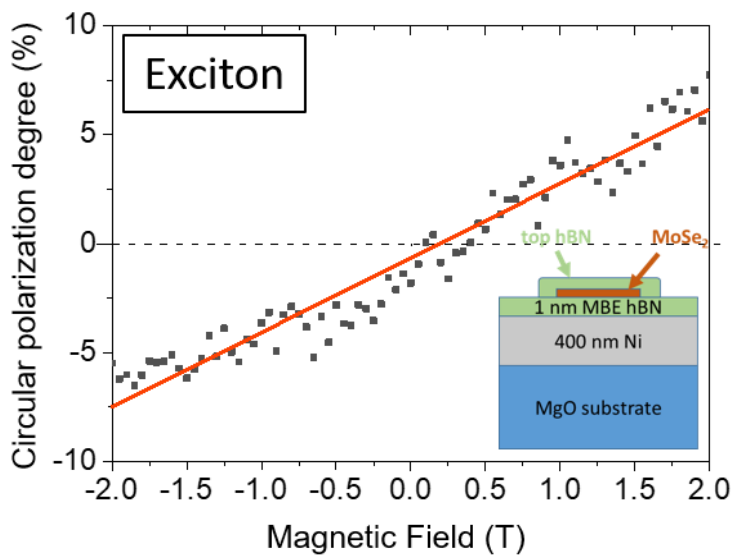

*Figure S3. Degree of circular polarization as a function of the external magnetic field for the neutral exciton in the Ni sample with the thinnest hBN barrier.*

3. <u>PL spectra as a function of temperature</u>

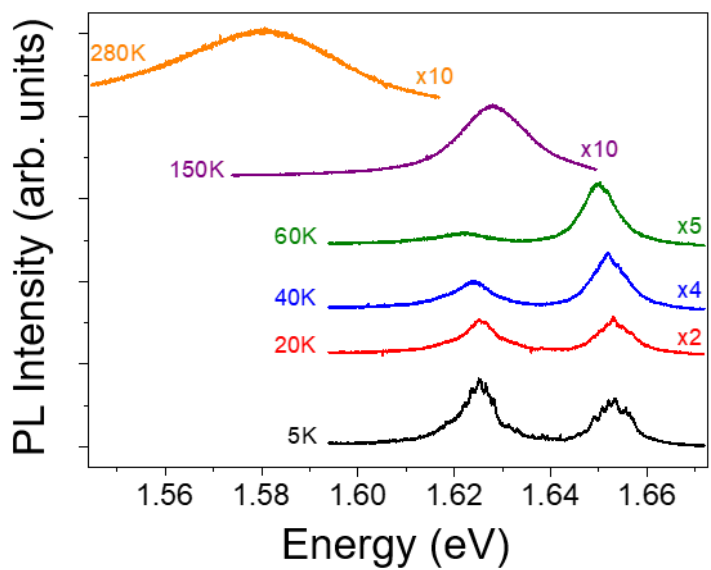

*Figure S4. PL spectra of the 1nm thick hBN sample as a function of temperature.*

4. <u>Estimation of the maximal leakage magnetic field in MoSe$_2$ due to the Ni layer (for a hBN thickness of 1 nm)</u>

In presence of roughness of the Ni layer, the maximal leakage magnetic field $B_{leak}$ is given by the formula[1]: $B_{leak} = (\mu_0 M_s/\pi)[1 -\exp(-2k\sigma)]\exp(-kt)$ where $\mu_0$ is the magnetic permeability in the vacuum, $M_s$ is the magnetization at saturation for Ni, t is the distance from the bottom of the magnetic layer, σ characterizes the average amplitude of the fluctuations of Ni surface roughness, and $k=\sqrt{2}\pi/L$, with L the correlation length that characterizes the spatial dependence of the fluctuations of Ni surface roughness.

Thanks to the VSM measurement, we can estimate that $M_s$ is about 4.3 10$^5$ A/m. Thanks to an atomic force microscopy (AFM) profile of Ni[2], we can estimate in a crude approximation that σ is about 0.3 nm under our spot size, and L is in between 20nm-120 nm. This leads for t=1nm to a leakage magnetic field in between about 3.5 mT and 17mT, far below the additional magnetic field (about 350 mT) that would be able to create an additional circular polarization Δ of 4%. We can thus rule out the leakage magnetic field as a possible source of Δ.